\documentclass[prd,nofootinbib,preprint,aps,amsfonts]{revtex4}
\def\lsim{\mathrel{\rlap{\lower4pt\hbox{\hskip1pt$\sim$}}
    \raise1pt\hbox{$<$}}}      
\def\gsim{\mathrel{\rlap{\lower4pt\hbox{\hskip1pt$\sim$}}
    \raise1pt\hbox{$>$}}}      
\begin{document}
	\title{Fermion family number and the  Z-Z$^{\prime}$ mixing in the 3-3-1 model with right-handed neutrinos }
	
	\author{D. Cogollo\footnote{E-mail: diegocogollo@fisica.ufpb.br}, H. Diniz\footnote{E-mail: hermes@fisica.ufpb.br}, C. A. de S. Pires\footnote{E-mail: cpires@fisica.ufpb.br}, P. S. Rodrigues da Silva\footnote{E-mail: psilva@fisica.ufpb.br}}
	\affiliation{Departamento de F\'{\i}sica, Universidade Federal da
	Para\'{\i}ba, Caixa Postal 5008, 58059-970, Jo\~ao Pessoa - PB,
	Brazil.}
	
	
	\begin{abstract}
	\vspace*{0.5cm}
	Theoretical consistency of the  3-3-1 model with right-handed neutrinos demands that  the number of family of  fermions be exactly equal to three.  In this brief report we show that such theoretical requirement results in a clean and severe bound on the Z-Z$^{\prime}$ mixing angle: $-3,979\times 10^{-3}<\phi<1,309\times 10^{-4}\,\,\,\,\,\mbox{with 90\% CL}$.
	\end{abstract}
	\maketitle
	
	%
	\section{Introduction}
	Although the standard model (SM) is not able to give any explanation to the problem of family replication, its neutral gauge boson sector 
is experimentally consistent with three families of fermions in nature.
	
To understand this, remember that, in the SM the invisible decay width of  its neutral gauge boson is solely due to light neutrinos. This allows us to define  the number of light neutrino species ($N_\nu$) through the ratio  
	\begin{eqnarray}
		N_\nu=\frac{\Gamma_{inv}}{(\Gamma_{\overline{\nu}\nu})_{SM}},
	\label{firstN}
\end{eqnarray}
where $\Gamma_{inv}$ is the measured invisible partial $Z$ decay width and $(\Gamma_{\overline{\nu}\nu})_{SM}$  is the SM expectation for the $Z$ partial decay width for each neutrino specie. The measurement of $\Gamma_{inv}$ is done by subtracting the measured visible partial $Z$ width into  quarks and charged leptons from the total $Z$ decay width. The $\Gamma_{inv}$  is assumed to be due to $N_\nu$ light neutrino species, each contributing with $\Gamma_{\overline{\nu}\nu}$, as given by the relation above. In order to reduce the model uncertainties, the relation $(\frac{\Gamma_{\bar l l}}{\Gamma_{\overline{\nu}\nu}})_{SM}$ is used instead of $(\Gamma_{\overline{\nu}\nu})_{SM}$. In this case, the expression to $N_\nu$  takes the form
\begin{eqnarray}
	N_\nu=\frac{\Gamma_{inv}}{\Gamma_{\bar l l}}(\frac{\Gamma_{\bar l l}}{\Gamma_{\overline{\nu}\nu}})_{SM}=R^{exp}(\frac{\Gamma_{\bar l l}}{\Gamma_{\overline{\nu}\nu}})_{SM}.
	\label{NSM}
\end{eqnarray}
The bounds for $N_{\nu}$ involves the experimental value for the ratio $R^{exp}=\frac{\Gamma_{inv}}{\Gamma_{\bar l l}}=5.942\pm0.016$\cite{Ndefinition} and the SM expectation to the ratio $(\frac{\Gamma_{\bar l l}}{\Gamma_{\overline{\nu}\nu}})_{SM}$ , where $(\frac{\Gamma_{\overline{\nu}\nu}}{\Gamma_{\bar l l}})_{SM}=1.991\pm0.001$\cite{Ndefinition}. In other words, the value of $N_\nu$  is model dependent. In the SM, the combined results from the four LEP experiments gives $N_\nu=2.984\pm 0.008$\cite{Ndefinition}. Notice that, in the SM, the value of $N_\nu$ is not necessarily an integer number. Moreover,  as in the SM the number of light neutrino species  coincides with the number of fermion families, thus we can conclude that the SM is bounded to have three families of fermions.

In extensions of the SM where the contributions to $\Gamma_{inv}$ are solely due to light neutrinos, the bound for the number of light neutrinos species is  given by the expression in Eq. (\ref{NSM}), with the correct modification of the SM expectation for the ratio $\frac{\Gamma_{\bar l l}}{\Gamma_{\overline{\nu}\nu}}$ by the corresponding expression for this ratio in the considered model. In the particular case of gauge extensions of the SM, the neutral gauge boson that will play the role of the standard one is now called  $Z_1$ which is a mix of $Z$  and $Z^{\prime}$ neutral gauge bosons. As a result $\frac{\Gamma_{\bar l l}}{\Gamma_{\overline{\nu}\nu}}$ will present a dependence on the Z-Z$^{\prime}$ mixing angle. Thus the bounds for the number of light neutrino species in this case will depend on this mixing angle. Analysis of the behaviour of $N_\nu$ with the Z-Z$^{\prime}$ mixing angle in the left-right model were done in  Refs.~\cite{miranda,perez}.  

In this work we are interested in the 3-3-1 model with right handed neutrinos (331RH$\nu$)\cite{pleitez,footpp}. In it the left-handed as well as the right-handed neutrinos are naturally light~\cite{light} in such a way that $\Gamma_{inv}$ is solely due to light neutrinos. 

Following the arguments above,  the corresponding expression for the number of light neutrino species in the 331RH$\nu$ is given by
\begin{eqnarray}
	N_\nu=R^{exp}(\frac{\Gamma_{\bar l l}}{\Gamma_{\overline{\nu}\nu}})_{331},
	\label{N331}
\end{eqnarray}
where $(\frac{\Gamma_{\bar l l}}{\Gamma_{\overline{\nu}\nu}})_{331}$ is the 331RH$\nu$ expectation for this ratio. 

Now comes the main point of this work. In way that the 331RH$\nu$ presents theoretical consistency, the number of family of fermions  must be exactly equal to three\cite{footpp,stu}. Moreover, notice that  the number of family of fermions in the 331RH$\nu$ coincides with the number of species of light neutrinos. This means that, in the 331RH$\nu$,  $N_\nu=3$, which is a true prediction of the model. Thus, the idea of this brief report is to use Eq. (\ref{N331}), but now having in mind that $N_\nu=3$, to extract bounds on the Z-Z$^{\prime}$ mixing angle in the 331RH$\nu$. 
\section{The Model}

We start presenting some aspects of the model concerning its leptonic sector. The 3-3-1 model with right-handed neutrinos\cite{pleitez,footpp} is one of the possible models allowed by the $SU(3)_C \otimes SU(3)_L \otimes U(1)$  gauge symmetry where  the leptons are distributed in the following representation content,
\begin{eqnarray}
f_{aL} = \left (
\begin{array}{c}
\nu_{aL} \\
e_{aL} \\
(\nu_{aR})^{c}
\end{array}
\right )\sim(1\,,\,3\,,\,-1/3)\,,\,\,\,e_{aR}\,\sim(1,1,-1),
 \end{eqnarray}
where $a = 1,\,2,\, 3$ refers to the three generations. Note that the right-handed neutrinos compose, with the left-handed ones,  the same triplet.

The masses of the charged leptons are generated by the scalar triplet $\rho \sim ({\bf 1}\,,\,{\bf 3}\,,\,2/3)$, with the vacuum $\langle \rho \rangle_0 = \frac{1}{\sqrt{2}}\left (0 \,\,,\,\,v_\rho \,\,,\,\, 0 \right )^T$, while the masses of the neutrinos are generated by the scalar triplets $\eta \sim ({\bf 1}\,,\,{\bf 3}\,,\,-1/3)$ and $\chi \sim ({\bf 1}\,,\,{\bf 3}\,,\,-1/3)$ with the corresponding vacua $ \langle \eta \rangle_0 = \frac{1}{\sqrt{2}}\left ( v_\eta \,\,,\,\,0\,\,,\,\, 0 \right )^T$  and $ \langle \chi \rangle_0 = \frac{1}{\sqrt{2}}\left (0\,\,,\,\, 0\,\,,\,\,v_{\chi^{\prime}}
\right )^T$.

In this point, I would like to call the attention to the fact that recently MiniBooNE collaboration released a report which refutes the LSND signal of mixing among left-handed and right-handed neutrinos  with 98\% CL\cite{miniboonereport}. It is interesting to remember that LSND signal\cite{LSNDsignal1,LSNDsignal2,LSNDsignal3}  was the unique experiment claiming a mixing among left-handed and right-handed light neutrinos.  After the release of the  MiniBooNE report, if we still consider this kind of mixing among left-handed and right-handed neutrinos, we must take into account the  cosmological bound on the angle $\theta$, which is about $\sin^2(2\theta)\approx 10^{-9}$\cite{cosmobounds1,cosmobounds2} for right-handed neutrinos of mass around KeV. In view of such tiny value for this mixing angle, we think reasonable to neglect it  in this work. For this, we make use of the following set of discrete symmetries over leptons and scalars,
\begin{eqnarray}
	\left( \chi\,,\,\rho\,,e_{aR}\right) \rightarrow -\left( \chi\,,\,\rho\,,e_{aR}\right),
	\label{discretesymmetryI}
\end{eqnarray}
which, besides avoiding mixing among left-handed and right-handed neutrinos, they also  avoid undesired terms in the Yukawa sector.
In order to see this, note that, with this, the only  Yukawa interactions involving leptons is $G_{ab}\bar f_{aL} \rho e_{bR}$. In this model left-handed as well as right-handed neutrinos gain masses through effective dimension-five operators. Following Ref. \cite{light}, and obeying the discrete set of symmetries above, the only effective operators we can build with these scalars and the  matter  content are these,
\begin{eqnarray}
{\cal L}_{M_L}&&=	\frac{g_{ab}}{\Lambda}\left( \overline{(f_{aL})^C} \eta^* \right)\left( \eta^{\dagger}f_{bL} \right)+ \frac{h_{ab}}{\Lambda}\left( \overline{(f_{aL})^C} \chi^* \right)\left( \chi^{\dagger}f_{bL} \right)+\mbox{H.c}.
	\label{5Doperators}
\end{eqnarray}	
According to these operators, when $\eta^0$ and $\chi^{\prime 0}$ develop VEV, $v_\eta$ and $v_{\chi^{\prime}}$, respectively, the left-handed and right-handed neutrinos both develop the following Majorana mass terms
\begin{eqnarray}
m^L_{ab}\overline{(\nu_{aL})^C} \nu_{bL}\,\,\,\,;\,\,\,\,m^R_{ab}\overline{(\nu_{aR})^C} \nu_{bR}.
	\label{massnuLR}
\end{eqnarray}
where $m^L_{ab}=\frac{g_{ab}v^2_\eta}{\Lambda}$  and $m^R_{ab}=\frac{h_{ab}v^2_{\chi^{\prime}}}{\Lambda}$. To have an idea of the order of magnitude of the masses of the left-handed and right-handed neutrinos, let us assume that  $\Lambda =10^{14}$ GeV (GUT scale), $v_\eta =10^2$GeV(the electroweak scale), $v_{\chi^{\prime}}=10^3$GeV(the 331 scale) and consider $g,h={\cal O}(1)$. With this  we have  $m^L \approx 0.1$eV  and $m^R \approx 10$eV, which are light neutrinos.

The 331RH$\nu$ disposes of two neutral gauge bosons Z and Z$^{\prime}$. However, after the breaking of the $SU(3)_C \times SU(3)_L \times U(1)_N$  symmetry to the $SU(3)_C \times U(1)_{e}$ one, Z get mixed with  Z$^{\prime}$, forming the physical neutral gauge bosons $Z_1$  and $Z_2$\cite{z-zp},
\begin{eqnarray}
&&Z_1=Zc_\phi-Z^{\prime}s_\phi, \nonumber \\
&&Z_2=Zs_\phi+Z^{\prime}c_\phi.
\label{mix}
\end{eqnarray}
In view of this mixing, $Z_1$ turns to play the role of the standard neutral gauge boson, while $Z_2$ is the heavy one.

The right-handed neutrinos are singlets by the $SU(2)_L \otimes U(1)_Y$ symmetry. Thus, in principle,  they should interact with the neutral gauge bosons Z$^{\prime}$ only.  However, due to the Z-Z$^{\prime}$ mixing, these neutrinos also interact with $Z_1$ and then  contribute to its invisible decay width.

In regard to the leptonic neutral current associated with  $Z_1$, we have that the neutral gauge boson interactions involving $Z_1$ and the charged leptons can be expressed by the following lagrangian term,
\begin{eqnarray}
	{\cal{L}}^{NC}_{l}=\frac{g}{2c_{w}}\overline{l}\gamma^{\mu}a\left\{g_{V}-g_{A}\gamma_{5}\right\}lZ^{1}_{\mu},
	\label{lNC}
\end{eqnarray}
where $a=(c_{\phi}-\frac{1}{\sqrt{3-4s^{2}_{w}}}s_{\phi})$,\, \,$g_{V}=(\frac{1}{2}-2s^{2}_{w})$\, and \, $g_{A}=\frac{1}{2}$, with $\phi$ being the Z-Z$^{\prime}$ mixing angle and $w$ is the Weinberg angle, while the neutral gauge boson interactions involving $Z_1$ and the left-handed and right-handed neutrinos are given by,
\begin{eqnarray}
{\cal{L}}^{NC}_{\nu}=-\frac{g}{2c_{w}}\left[\overline{\nu}^{i}_{L}\gamma^{\mu}(c_{\phi}+\frac{1-2s^{2}_{w}}{\sqrt{3-4S^{2}_{w}}}s_{\phi})\nu^{i}_{L}\right]Z^{1}_{\mu}-\frac{g}{2c_{w}}\left[\overline{\nu}^{i}_{R}\gamma^{\mu}(\frac{2c_{w}^{2}}{\sqrt{3-4s^{2}_{w}}}s_{\phi})\nu^{i}_{R}\right]Z^{1}_{\mu},
	\label{Z1nu}
\end{eqnarray}
where $i=1,2,3$. Note that when $\phi \rightarrow 0$, the neutral current above is identified with the standard one involving left-handed neutrinos and the neutral gauge boson Z.
\section{$N_\nu$ bounds on $\phi$}
As we said above, theoretical consistency of the 331RH$\nu$ requires three generations of fermions. Let us now obtain the phenomenological bounds on the Z-Z$^{\prime}$ mixing angle according to this requirement. 

Perceive that the framework of the model ensures that three families of fermions imply necessarily in three types of neutrino species. Moreover, see also that, in the model, only neutrinos contribute to the invisible $Z_1$ decay width. With all this in mind, we can say that  the  number of light neutrino species  $N_\nu$ is given by the relation,
\begin{eqnarray}
	N_\nu=\frac{\Gamma_{inv}}{(\Gamma_{\overline{\nu}\nu})_{331}}.
	\label{firstN331}
\end{eqnarray}
In the model as the left-handed as well as  the right-handed neutrinos are light ones. Thus  the $\Gamma_{inv}$   in Eq. (\ref{firstN331}) is assumed  to be due to $N_\nu$ light left-handed and right-handed neutrinos, in other words, 
\begin{eqnarray}
	(\Gamma_{\overline{\nu}\nu})_{331}=\Gamma_{\overline{\nu_L}\nu_L}+\Gamma_{\overline{\nu_R}\nu_R},
	\label{nudecay331}
\end{eqnarray}
each contributing to $(\Gamma_{\overline{\nu}\nu})_{331}$ as given below,
\begin{eqnarray}
&& \Gamma_{\overline{\nu_L}\nu_L}= \frac{M^{3}_{Z_1}G_{F}}{12\sqrt{2}\pi}\left(c_{\phi}+\frac{1-2s^{2}_{w}}{\sqrt{3-4s^{2}_{w}}}s_{\phi}\right)^2  \\ \nonumber && 
\Gamma_{\overline{\nu_R}\nu_R}=\frac{M^{3}_{Z_1}G_{F}}{12\sqrt{2}\pi}\left(\frac{4c^{4}_{w}}{3-4s^{2}_{w}}s^{2}_{\phi} \right),
\label{left-rightwidth}	
\end{eqnarray}
according to the interactions in Eq. (\ref{Z1nu}).

Following the same procedure as for the case in the SM, we should have 
\begin{eqnarray}
	N_\nu=R^{exp}(\frac{\Gamma_{\bar l l}}{\Gamma_{\overline{\nu}\nu}})_{331},
\label{Nnu331model}
\end{eqnarray}
where $(\Gamma_{\bar l l})_{331}$ is the 331RH$\nu$ expectation for the $Z_1$ partial decay width for each pair of chargeded leptons, and according to the interaction in Eq. (\ref{lNC}), we have
\begin{eqnarray}
(\Gamma_{\overline{l}l})_{331}=\frac{M^{3}_{Z_1}G_{F}}{6\sqrt{2}\pi}(c_{\phi}-\frac{1}{\sqrt{3-4s^{2}_{w}}}s_{\phi})^{2}\left[(\frac{1}{2}-2s^{2}_{w})^{2}+\frac{1}{4}\right].
\label{widthll}
\end{eqnarray}

On substituting the expressions to $(\Gamma_{\bar l l})_{331}$  and  $(\Gamma_{\overline{\nu}\nu})_{331}$ given above in Eq. (\ref{Nnu331model}), we obtain
\begin{eqnarray}
	N_\nu=R^{exp}\frac{2\left((\frac{1}{2}-2s^{2}_{w})^{2}+\frac{1}{4}\right)(c_{\phi}-\frac{1}{\sqrt{3-4s^{2}_{w}}}s_{\phi})^{2}}{ (c_{\phi}+\frac{1-2s^{2}_{w}}{\sqrt{3-4s^{2}_{w}}}s_{\phi})^{2}+\frac{4c^{2}_{w}}{3-4s^{2}_{w}}s^{2}_{\phi}}.
	\label{331modelratio}	
\end{eqnarray}
Using the values $s^{2}_{w}=0,23122$\cite{Ndefinition} and $R^{exp}=\frac{\Gamma_{inv}}{\Gamma_{\bar l l}}=5.942\pm0.016$\cite{Ndefinition}, the requirement that $N_\nu$ is exactly equal to three results in the following range of allowed values for the mixing angle $\phi$,
\begin{eqnarray}
-3,979\times 10^{-3}<\phi<1,309\times 10^{-4}\,\,\,\,\,\mbox{with 90\% CL}.
\label{bounds}
\end{eqnarray}
This is a severe bound and, differently from other bounds\cite{otherbounds1,otherbounds2}, it does not depend on other parameters of the model, as the mass of the neutral gauge boson $Z_2$. Besides, it is a very clean bound in comparison to bounds that involve quarks. This is so because the Z-Z$^{\prime}$ mixing induces process that presents FCNC with the standard quarks. We think that all this  turns the bound we got here on Z-Z$^{\prime}$ mixing angle a very distinctive bound.

\section{concluding remarks}
The main goal of this work was to extract bounds on the Z-Z$^{\prime}$ mixing angle from  the fact that theoretical consistency of the  331RH$\nu$ demands that the number of family of fermions in nature must be exactly equal to three. As argued above, three family of fermions  imply that there are three species of light neutrinos in the model. On the other side, the number of light neutrino species is strongly related to the invisible $Z_1$ decay width. Thus we made use of the fact that in the model $N_\nu$ must be exactly equal to three to extract bounds on the Z-Z$^{\prime}$ mixing angle. We obtained the bound $-3,979\times 10^{-3}<\phi<1,309\times 10^{-4}\,\,\,\,\,\mbox{with 90\% CL}$. This is a very severe and interesting bound. It is interesting by two reasons. First, it is a very clean bound, in the sense that  the bound we got does not present any dependence on the mass of the heavy neutral gauge boson $Z_2$ neither involves the hadronic sector, which jeopardize any attempt of getting  bounds on  $\phi$ because such sector presents FCNC with the $Z_1$ gauge boson. The second reason is that it was derived from first principles, i.e., from the requirement that the model be theoretically consistent, which means that $N_\nu=3$ was a prediction of the model differently of SM where it is allowed to assume, in principle, any values.

\acknowledgments
This work was supported by Conselho Nacional de Pesquisa e
Desenvolvimento Cient\'{i}fico- CNPq(HD,CASP,PSRS) and Coordena\c c\~ao de Aperfei\c coamento de Pessoal de N\'{i}vel Superior - capes(DC).


%

\end{document}